\documentclass[conference]{IEEEtran}
\IEEEoverridecommandlockouts

\usepackage{cite}
\usepackage{amsmath,amssymb,amsfonts}
\usepackage{algorithmic}
\usepackage{graphicx}
\usepackage{textcomp}
\usepackage{xcolor}
\usepackage{url}
\usepackage{xspace}
\usepackage{authblk}
\usepackage{pdfpages}

\def\BibTeX{{\rm B\kern-.05em{\sc i\kern-.025em b}\kern-.08em
    T\kern-.1667em\lower.7ex\hbox{E}\kern-.125emX}}
\begin{document}

\title{Opening Portals:\\Relay Attacks on Apple's AirTags}
\title{A Relay a Day Keeps the AirTag Away:\\Practical Relay Attacks on Apple’s AirTags}

\makeatletter
\renewcommand\AB@affilsepx{, \protect\Affilfont}
\setlength{\affilsep}{0.5em}   %
\makeatother

\author[1]{Gabriel K. Gegenhuber}
\author[2]{Sebastian Strobl}
\author[1]{Leonid Liadveikin}
\author[1]{Florian Holzbauer}
\affil[1]{University of Vienna, Faculty of Computer Science}
\affil[2]{SBA~Research}

\maketitle

\def\findmy{\textsl{Find~My}\xspace}

\begin{abstract}

Apple AirTags use Apple's \findmy network: when nearby iDevices detect a lost tag, they anonymously forward an encrypted location report to Apple, which the tag's owner can then fetch to locate the item.
That encryption protects privacy ---neither the finder nor Apple learns the owner's identity--- but it also prevents Apple from validating the correctness of received reports.

We show that this design weakness can be exploited: using a relay attack, we can inject manipulated location reports so the \findmy service reports a false position for a lost AirTag.
The same technique can be used to deny recovery of a targeted tag (a focused DoS), since the owner is misled about its whereabouts.

\end{abstract}


\section{Introduction}

With Apple's \findmy infrastructure~\cite{2019_Apple_FM}, users can locate lost iDevices and ---since the 2021 introduction of AirTags~\cite{2022_Apple_A,2021_Apple_AIA}--- everyday objects as well.
AirTags' small form factor and long battery life (up to one year on a coin cell) make them convenient for tagging keys, bags, or bicycles~\cite{2022_Apple_A}, but they also enable covert misuse: AirTags can be concealed on a person or in personal belongings and used to track individuals without their consent~\cite{heinrich2024please, gerhardt2025airtag, turk2023can}.
While both Apple and Android provide countermeasures to reduce unwanted tracking, studies reveal that these protections can suffer from substantial delays or outright failures, during which victims may unknowingly leak sensitive location information~\cite{2022_Heinrich_APAU, turk2023can}.

The \findmy network relies on hundreds of millions of participating iDevices to observe nearby lost items and forward encrypted geolocation reports to Apple; the item's owner can then retrieve these reports to determine its location~\cite{2021_Heinrich_WCFM, 2019_Apple_FM}.
While this design preserves privacy (the finder and owner remain anonymous), the encryption also prevents Apple from validating the integrity of submitted reports.
In this paper, we exploit that gap: using a practical relay technique, we capture and rebroadcast an AirTag's Bluetooth Low Energy (BLE) signal at a different place, causing nearby iDevices to generate and upload falsified location reports.
As a result, the \findmy app may display an incorrect position for the lost tag, effectively producing a targeted Denial-of-Service (DoS) against recovery.

We validate the attack under realistic conditions
and perform empirical characterization of AirTag behavior ---including advertisement rotation and the usable lifetime of captured signals for replay/relay--- to quantify the attack surface and limits.
Our results expose a fundamental trade-off between privacy and integrity in the \findmy design and motivate countermeasures that preserve user privacy while ensuring the authenticity of location reports.

\section{Methodology and Implementation}
We divided our study into three stages: (i) passive observation, (ii) emulation and relaying, and (iii) exploitation.
Each stage built on the previous one to progressively characterize AirTag behavior and validate the relay attack under realistic conditions.

\begin{figure}[b!]
  \centering
  \includegraphics[width=1\columnwidth]{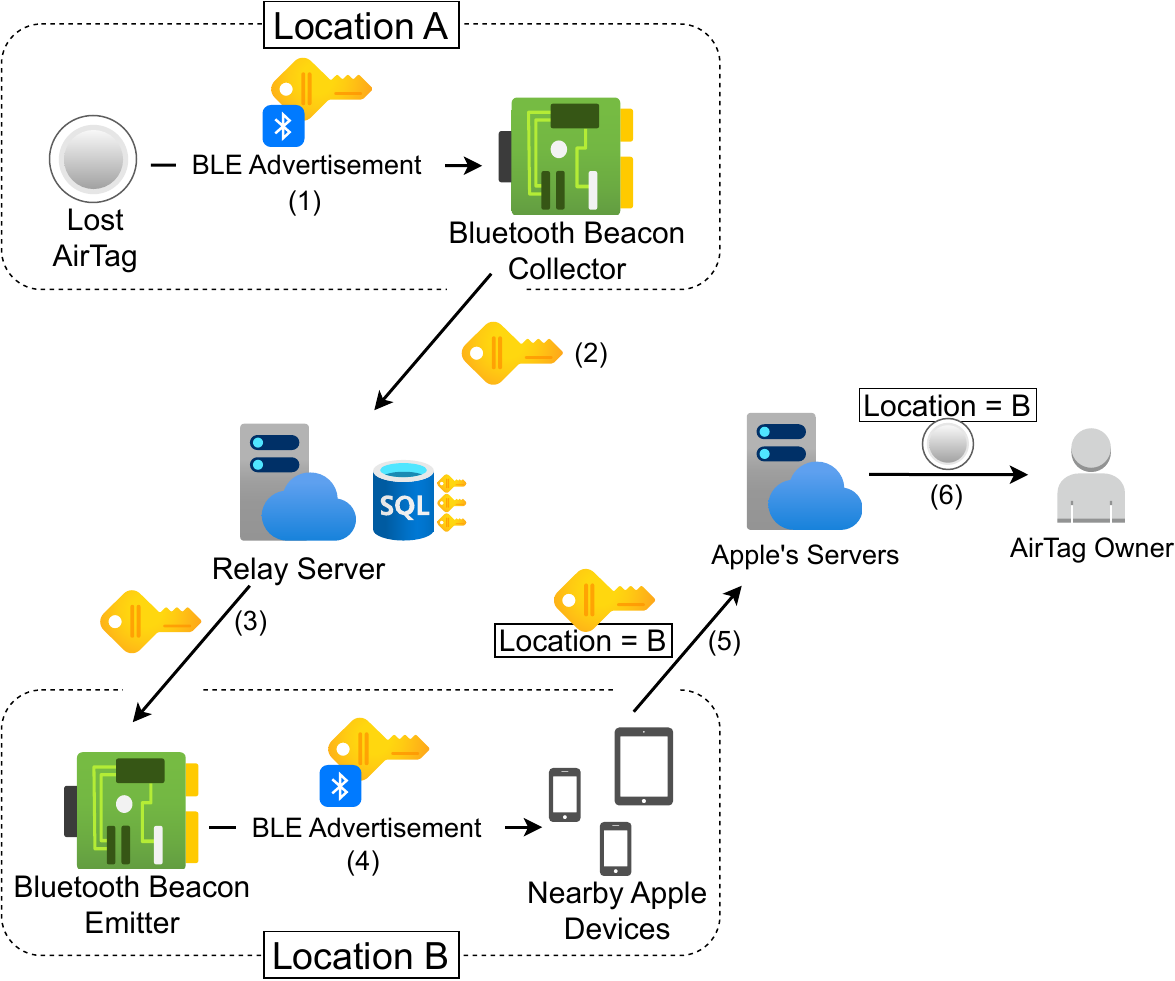}
  \caption{Relaying an AirTag's BLE advertisments over the Internet injects false location reports into the \findmy system.}
  \label{fig:airtag_relay}
\end{figure}

All experiments were performed using our own devices and infrastructure and in accordance with applicable institutional research and ethics guidelines.

\subsection{Passive Observation and Beacon Collection}
To isolate signals from our test AirTag and minimize environmental noise, we conducted experiments in a controlled environment and continuously captured BLE advertisements.
Using established open-source tooling (AirGuard~\cite{heinrich2022airguard}, OpenHaystack~\cite{burg2022openhaystack}), we developed lightweight BLE \emph{collector} scripts for Android, Linux, and ESP32 platforms to record raw advertisements, timestamps, RSSI, and relevant advertisement fields.
From these captures, we identified the packets belonging to our device, observed state transitions between \emph{connected} (paired to the owner’s phone) and \emph{lost} modes, and confirmed regular key rotation behavior (for AirTags, once every 24h) reported in prior work~\cite{2021_Heinrich_WCFM, 2021_Mayberry_WTtT}.

\subsection{Emulation and Relay}
Second, we implemented the ability to emulate and rebroadcast captured beacons.
After collecting advertisements from our test AirTag, we removed the battery from the original tracker device and used custom (non-official) hard- and software to retransmit the recorded packets.
AirTags embed parts of their public key material in the BLE MAC address; because changing the BLE MAC on unrooted Android devices is not possible, we implemented BLE \emph{emitter} code for Linux and ESP32 targets that permit MAC manipulation.

We validated successful emulation by observing that the owner's \findmy client retrieved location reports corresponding to the emulated broadcasts.
To demonstrate remote relaying, we rebroadcast the captured beacon at a remote location (i.e., a Raspberry Pi located at a research institute in another country); iDevices near that site generated and uploaded geolocation reports for the emulated tag, which our owner device was able to fetch, confirming that relayed signals can produce falsified location reports.

\subsection{Testing Corner Cases and Exploitation}
Finally, we evaluated the practical limits and abuse potential of captured and retransmitted beacons.
Our goal was to determine the usable lifetime of a captured advertisement before it can no longer be used to create false reports.
To facilitate our experiments, we implemented a relay server that ingests reports from collector devices, timestamps first occurrence, stores raw advertisements, and provides controlled replay scheduling to emitter nodes.
Figure~\ref{fig:airtag_relay} shows our experimental setup.

\section{Results}
\begin{figure}[t]
  \centering
  \includegraphics[width=.8\columnwidth]{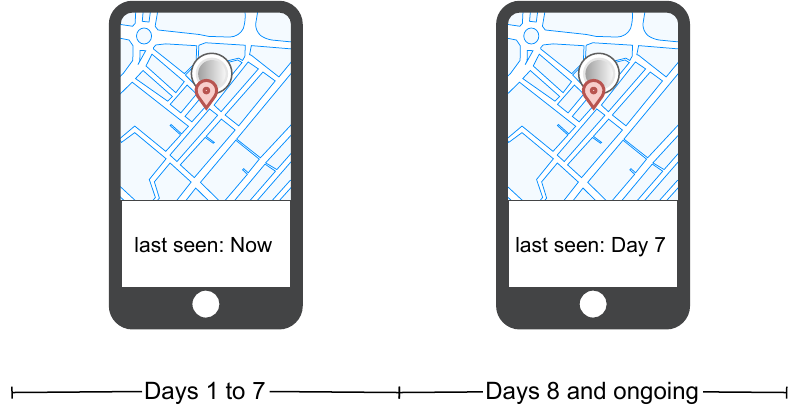}
  \caption{While, key rotations on the AirTag invalidate previous keys (for cloud reports), historic beacons can still be used to mislead the tag owner via local BLE advertisments.
  When surpressing key rotations on the AirTag (e.g., by removing the battery), captured BLE beacons can be replayed for seven days before the \findmy app flags them as outdated. }
  \label{fig:reuse_ad}
\end{figure}

Our findings show, that the \findmy app differentiates between two data sources:
\begin{itemize}
    \item \textbf{Cloud Reports}: Location reports that come from the \findmy cloud are only considered when they use the most-recent public key.
    Thus, key rotations on the AirTag invalidate all previous keys.
    When an AirTag's current BLE beacons were continuously relayed to a remote location, the \findmy app alternated between the legitimate and relayed positions, causing jumps between the two competing locations.
    
    \item \textbf{Local BLE Beacons}: When the AirTag appears within the BLE range of the owner device, the \findmy app assumes that it is close and therefore ignores any data that is reported via the cloud.
    Overriding cloud reports with local advertisements also worked with previously recorded i.e., \textit{historic} beacons. 
\end{itemize}
The \findmy app also differentiates between recent and outdated reports, as show in Figure~\ref{fig:reuse_ad}.

\section{Conclusion and Future Work}
Our results demonstrate that the design of Apple's \findmy protocol is vulnerable to practical relay attacks. Beyond highlighting this attack vector to the security community, future work will examine whether relaying could be repurposed as an active countermeasure against stalking by obfuscating or disrupting unwanted tracking. 
Therefore, we aim to investigate which algorithms (e.g., averaging) are used to reconcile multiple recent location reports.
Finally, we plan to broaden our study to include additional tracker ecosystems (e.g., Chipolo, Samsung SmartTag, Tile) to assess whether similar weaknesses exist across competing platforms.

\bibliographystyle{plain}
\bibliography{sebastian, bibliography}

\pagebreak


\section*{Supplementary material}

# A Relay a Day Keeps the AirTag Away:
## Practical Relay Attacks on Apple's AirTags

Gabriel K. Gegenhuber, Leonid Liadveikin, Florian Holzbauer (University of Vienna), Sebastian Strobl (SBA Research)

## Background & Motivation

Apple's *Find My* network enables users to locate lost devices and AirTags by **crowdsourcing** location data from nearby iPhones.

▶ AirTags **broadcast** Bluetooth Low Energy (BLE) beacons containing their **public key**.
▶ Nearby **iDevices (e.g., iPhones)** detect these and **upload encrypted location reports** to Apple.
▶ Encryption preserves user privacy, but also prevents Apple from verifying whether the data is legitimate.

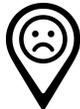

**Misuse:** AirTags can facilitate **unwanted tracking or stalking**.
▶ Both Android and iOS implement anti-stalking notifications and detection features.
▶ Existing solutions can suffer from **substantial delays or failures**.
▶ Victims are often alerted only after their location is exposed.

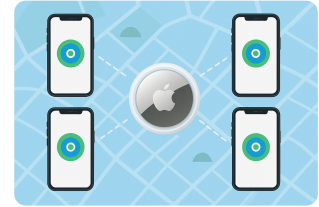

## Research Objective

To demonstrate and analyze a **practical relay attack** that can:

▶ Inject **false locations** into Apple's Find My network.
▶ **Mislead** owners or **deny recovery** of a tag (targeted DoS).
▶ Evaluate whether this behavior could also be leveraged to **protect against unwanted tracking**.

## Methodology

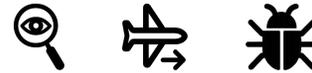

(i) Passive Observation
▶ Capturing BLE advertisements from our **own AirTag** in a **controlled environment**.
▶ Observing key roatations (every 24 h) and state transitions (paired vs. lost state).

(ii) Emulation & Relay
▶ Confirming relay attack: rebroadcasted BLE beacons generate **false location reports**.
▶ Verified attack over **long-distances** (between countries).

(iii) Controlled Exploitation
▶ Evaluating practical limits and abuse potential of relaying.
▶ Measuring the **usable lifetime** of captured beacons.

## Experimental Setup

Building upon existing open-source tools (AirGuard, OpenHaystack) and custom hardware (Android, Linux, ESP32), we created a **flexible relay infrastructure** for timestamping, storage, and scheduled replays.

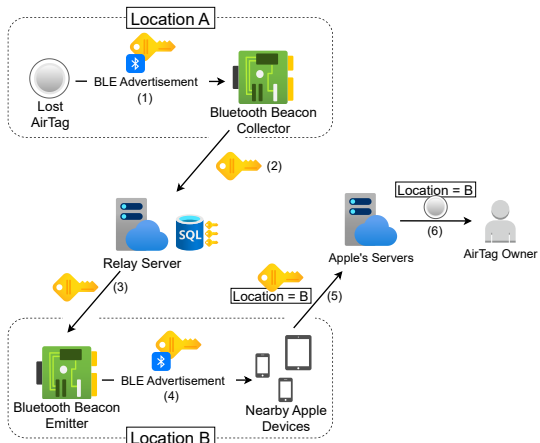

*Figure 1:* Relaying an AirTag's BLE advertisments over the Internet injects false location reports into the *Find My* system.

Using our tooling for recording and replaying AirTag BLE beacons, we show that Apple's *Find My* protocol is highly **vulnerable to practical relay attacks**. The findings underline a core limitation of Apple's privacy model: encryption protects identities but prevents the server from validating report integrity.

## Results & Observations

**Data Source Handling**
▶ **Local BLE beacons** are **preferred over cloud reports**, leading to DoS when owner device is in close proximity.

**Relay Effects**
▶ Continuous relaying to remote location leads to **jumping between** true and false **positions**.

**Beacon Lifetime**
▶ **Local BLE Reports:** Captured beacons remain valid for at least 7 days before being flagged as outdated.
▶ **Cloud Reports:** Key rotations on the AirTag invalidate all previous keys.

## Conclusion & Future Work

▶ **Relay attack feasible**, attackers can **mislead or block item recovery**.
▶ Relay mechanism can be **repurposed as a privacy shield**.
▶ **Extend analysis to other trackers**: Chipolo, Samsung SmartTag, Tile.
▶ Further explore ethical relay-based anti-stalking defenses.



\end{document}